# A new route for enantio-sensitive structure determination by photoelectron scattering on molecules in the gas phase

Kilian Fehre[1*], Nikolay M. Novikovskiy[2,3], Sven Grundmann[1], Gregor Kastirke[1], Sebastian Eckart[1], Florian Trinter[1,4], Jonas Rist[1], Alexander Hartung[1], Daniel Trabert[1], Christian Janke[1], Martin Pitzer[1], Stefan Zeller[1], Florian Wiegandt[1], Miriam Weller[1], Max Kircher[1], Giammarco Nalin[1], Max Hofmann[1], Lothar Ph. H. Schmidt[1], André Knie[2], Andreas Hans[2], Ltaief Ben Ltaief[2], Arno Ehresmann[2], Robert Berger[5], Hironobu Fukuzawa[6], Kiyoshi Ueda[6], Horst Schmidt-Böcking[1], Joshua B. Williams[7], Till Jahnke[8], Reinhard Dörner[1], Philipp V. Demekhin[2*], and Markus S. Schöffler[1*]

[1]Institut für Kernphysik, Goethe-Universität Frankfurt, Max-von-Laue-Straße 1, 60438 Frankfurt am Main, Germany
[2]Institut für Physik und CINSaT, Universität Kassel, Heinrich-Plett-Straße 40, 34132 Kassel, Germany
[3]Institute of Physics, Southern Federal University, 344090 Rostov-on-Don, Russia
[4]Molecular Physics, Fritz-Haber-Institut der Max-Planck-Gesellschaft, Faradayweg 4-6, 14195 Berlin, Germany
[5]Department of Chemistry, Philipps-Universität Marburg, Hans-Meerwein-Straße 4, 35032 Marburg, Germany
[6]Institute of Multidisciplinary Research for Advanced Materials, Tohoku University, Sendai 980-8577, Japan
[7]Department of Physics, University of Nevada, Reno, Nevada 89557, United States
[8]European XFEL GmbH, Holzkoppel 4, 22869 Schenefeld, Germany

*Correspondence to: fehre@atom.uni-frankfurt.de, demekhin@physik.uni-kassel.de, schoeffler@atom.uni-frankfurt.de

## Abstract

X-ray as well as electron diffraction are powerful tools for structure determination of molecules. Studies on randomly oriented molecules in the gas-phase address cases in which molecular crystals cannot be generated or the interaction-free molecular structure is to be addressed. Such studies usually yield partial geometrical information, such as interatomic distances. Here, we present a complementary approach, which allows obtaining insight to the structure, handedness and even detailed geometrical features of molecules in the gas phase. Our approach combines Coulomb explosion imaging, the information that is encoded in the molecular frame diffraction pattern of core-shell photoelectrons and ab initio computations. Using a loop-like analysis scheme we are able to deduce specific molecular coordinates with sensitivity even to the handedness of chiral molecules and the positions of individual atoms, as, e.g., protons.

## Introduction

During the last decade, the determination of the three-dimensional structure of molecules using electron crystallography developed into a complementary analysis method to the well-established X-ray crystallography *(1-3)*. In particular, for the structural investigation of micro- and nano-crystalline materials, where sufficiently large single crystals for X-ray diffraction cannot be obtained *(4)* or crystalline sponge approaches for X-ray diffraction *(5,6)* fail, electron diffraction is the method of choice *(7)*. It yields Ångstrom-resolution even when applied to large systems *(8)* or systems involving weak scatterers such as hydrogen atoms *(9)*. For cases in which molecular crystals cannot be obtained or the interaction-free molecular structure is to be addressed, corresponding electron scattering approaches on gas-phase molecules exist *(10,11)*. Such electron diffraction studies on randomly oriented molecules, however, can only provide information on interatomic distances, which is, in addition, challenging to extract in case of overlapping distance parameters. Furthermore, they do not reveal the handedness of chiral systems *(12)*. In order to overcome the drawback of a random orientation of the molecules under investigation, several experiments have been performed utilizing electron *(13,14)* and x-ray diffraction *(15)* in combination with sophisticated 2-dim. and 3-dim. laser alignment schemes.

An alternative approach employs electrons that are created from within the molecule as a probe. In these experiments, high-energetic single photons or strong-field laser pulses ionize isolated molecules in the gas phase. The emitted photoelectron is scattered by the molecular potential, yielding a very complex interference pattern, in which the structural information is encoded. Using table-top laser systems, laser-induced electron diffraction (LIED) has, for example, proven its capability determining internuclear distances with high accuracy for small molecules *(16-18)*. In general, photoelectron diffraction by molecules in the gas phase has been successfully applied for determining molecular constituents *(19)*, mapping bond lengths *(20,21)* and simple chemical reactions *(22)* on ultrafast timescales *(17)*. Until now, however, corresponding studies were restricted to linear *(17,19,22)* or mostly symmetric molecules *(23-25)* such as, e.g., CO, $CO_2$, $H_2O$, or $CH_4$.

Apart from measuring the electron diffraction pattern (in terms of an electron angular emission distribution), the key to electron diffraction experiments on molecules is the knowledge of each individual molecule's orientation in space *(26)*. A possible approach is to adsorb the molecule to a surface *(13)* or (as indicated above) to utilize weak laser pulses in various schemes (and in 2d or 3d arrangements) to orient *(27,28)* or align *(15,27-31)* the molecule. Single cycle pulses in the THz were also used in the past to orient/align molecules *(32)*. An alternative approach for detecting the molecule's spatial orientation is Coulomb explosion imaging *(33)*, which, in addition, provides structural information, as well. Here, molecular ions or molecules are rapidly charged up by foil-induced electron stripping *(34)*, multiple ionization by a short and strong laser pulse, or by photoionization and subsequent Auger decay (cascades) *(34)*. After the charge-up, the ionic fragments are driven apart rapidly by Coulomb repulsion. Intriguingly, if more than three molecular fragments are generated in the Coulomb explosion, the triple product of three of their momentum vectors allows for identifying whether a chiral molecule was right- or left-handed *(35)*. However, despite the absolute configuration of chiral molecules can be determined in principle using this method *(35,36)*, it has been restricted, so far, to small molecules with only a few atoms. So far, the largest molecule investigated using this approach was halothane consisting of eight atoms *(37)*, but just recently iodopyridine (11 atoms) has been addressed in an experiment *(38)*. Seribal et al. have shown in a simulation, that Coulomb Explosion Imaging in combination with a spatial orientation of the gas-phase target substance allows for retrieving the molecules' handedness even for system as large as camphor *(39)*. In detail, however, gathering structural information of larger molecules with the help of Coulomb explosion techniques faces yet multiple technical challenges. These are, for example, the initial generation of high charge states, the rapidly declining detection efficiency for the coincident detection of multiple molecular fragments *(40)*, and uncertainties in their correct m/q-assignment. In addition, the inversion of the measured momentum space information to position space is far from trivial as soon as the charge-up of the molecule and/or its fragmentation is governed by nuclear dynamics. Already when examining small molecules as $H_2O$, support from sophisticated theory is required for the interpretation of Coulomb explosion data, which in turn provides valuable details on the fragmentation dynamics and processes *(41)*.

In this article, we demonstrate in a proof-of-principle study, how to overcome these obstacles. With a combination of the concepts of Coulomb explosion imaging, photoelectron diffraction imaging and support from ab initio modelling, we developed a method for addressing isolated molecules in the gas phase to determine their structure and their handedness. Our scheme is applicable without the need for advanced laser-alignment schemes or elaborate abilities for detecting a multitude of ionic fragments in coincidence. We will show, furthermore, that our approach allows to determine even tiny details, as, for example, a slight displacement of a hydrogen atom in a methyloxirane molecule.

## Experimental method

In our study we target methyloxirane molecule and examine its ionic fragments (occurring after photoionization and subsequent Auger decay) and the angular distributions of the emitted photoelectrons. The measurements were performed employing Cold Target Recoil Ion Spectroscopy (COLTRIMS) which is a multi-coincidence momentum imaging technique *(42,43)*. In brief, ions and electrons created in the interaction of single photons from the Soleil synchrotron with single methyloxirane molecules were guided by electric and magnetic fields onto two time- and position-sensitive multichannel plate detectors. From the particles' positions of impact and times of flight the individual trajectories inside the COLTRIMS-spectrometer were reconstructed in an offline analysis of the data. This information yielded the particles' momenta and accordingly all derived observables as emission directions and kinetic energies. As a coincidence measurement has been performed, relative quantities are retrieved, as well, as for example relative emission angles. The experimental data were recorded at the same beam time as the data from a previous publication. Accordingly, the identical experimental setup was used and further details on the exact parameters of the COLTRIMS reaction microscope can found there *(44)*. The employed photon energy of $550$ eV addressed the O 1s-shell of the methyloxirane molecule leading to a photoelectron energy of about $11.5$ eV and we restricted our analysis to electrons with kinetic energy of $11.5 \pm 1.5$ eV in order to suppress possible background. Furthermore, we employed the aforementioned photon energy as photoelectrons of this specific kinetic energy showed a large chiral response and this electron energy is amenable for accurate calculations *(44)*. Several fragmentation pathways occur after the O-K-ionization of the molecule and subsequent Auger decay(s). As detailed below, we employ for our study cases where the molecule fragmented into at least three parts of which two are charged. The data presented in this study consists of a combination of the fragmentation channels $C_3H_6O \rightarrow C_2H_3^+ + CH_2^+ + OH^0 + 2e^-$ and $C_3H_6O \rightarrow C_2H_2^+ + CH_2^+ + H_2O^0 + 2e^-$. About $3 \cdot 10^6$ events were recorded for both enantiomers of the molecule and both light helicities. Two other breakup channels, namely $C_3H_6O \rightarrow C_2H_3^+ + CH_3^+ + O^0 + 2e^-$ and $C_3H_6O \rightarrow C_2H_2^+ + CH_3^+ + OH^0 + 2e^-$, cannot be used because of the following reasons: It turns out that for the latter fragmentation channel the $CH_3^+$ group stems from the methyl group of methyloxirane. This different fragmentation dynamics manifests itself in the fact that the measured momenta of the ions define a completely different molecular coordinate system. In addition, the interference patterns for these fragmentation channels are washed out. We suspect that this is due to a weaker correlation between measured ionic momenta and the molecular orientation at the instant of ionization due to complex fragmentation dynamics.

## Theoretical method

In order to extract information on the molecular geometry from the experimental data we performed a modeling of the electron diffraction pattern, i.e., of the molecular-frame angular emission distributions of the photoelectrons. The ionization transition amplitudes for the emission of O(1s) photoelectrons of the methyloxirane enantiomers were computed by using the single-center method and code *(45,46)* in the relaxed-core Hartree-Fock approximation, as described in our previous work on this molecule *(44)* (please see the supplementary information document of this reference for more details, as well).

The averaging of the molecular frame photoelectron angular emission distributions over all incident directions of the ionizing light, required for the present study, was performed analytically. The average differential cross section reads:

$$\frac{d\sigma(\theta,\varphi)}{d\Omega} = \sum_{LM} B_{LM} Y^{*}_{LM}(\theta,\varphi) \quad \text{with} \quad B_{LM} = \frac{1}{3}\sum_{\ell m}\sum_{\ell' m'}\sum_{k} i^{\ell+\ell'}(-1)^{\ell+m'} \times$$

$$\times \sqrt{\frac{(2\ell+1)(2\ell'+1)(2L+1)}{4\pi}} \begin{pmatrix} \ell & \ell' & L \\ 0 & 0 & 0 \end{pmatrix} \begin{pmatrix} \ell & \ell' & L \\ m & -m' & M \end{pmatrix} A_{\varepsilon\ell mk} A^{*}_{\varepsilon\ell' m' k}$$

Here, $\theta$ and $\varphi$ are the photoelectron emission angles in the frame of molecular reference, $Y_{LM}$ are spherical harmonics, and $A_{\varepsilon\ell mk}$ are the dipole transition amplitudes for the emission of the partial photoelectron waves with the angular momentum quantum numbers $\ell$ and $m$ via the absorption of a photon of polarization $k$, as defined in the frame of the molecule. Because of the mutual orthogonality of the Wigner rotational matrices (which transform the ionizing light of a given polarization from the laboratory to the molecular frame), the average molecular frame photoelectron angular distribution is independent of the polarization of the ionizing light that is used in the experiment.

## Results and Discussion

As mentioned above, we use synchrotron light to ionize the molecule by emission of a core electron. The emerging photoelectron is diffracted by the molecular potential and serves as a messenger providing the molecular structure information in its angular emission pattern. Molecules are in most cases unstable after the emission of a core electron. Typically, at least one more additional electron is released in an Auger decay process and subsequently the molecule fragments into charged and neutral pieces. It turns out, that the detection of the fragmentation direction (i.e., the momentum vectors) of two charged fragments of a breakup of the molecule into at least three pieces is sufficient to gather the information on the spatial orientation of the molecule, which is needed for evaluating the photoelectron interference pattern. Dictated by conservation of linear momentum, the momentum vectors of three molecular fragments lie within a plane (turquoise arrows in Fig. 1A). They can be employed to form a fragment coordinate frame (X,Y,Z) in Fig. 1A, which was built as follows: $\vec{p}_{C_2H_3^+}$($\vec{p}_{C_2H_2^+}$ respectively) points in the direction of the Y-axis, $\vec{p}_{C_2H_3^+} \times \vec{p}_{CH_2^+}$ ($\vec{p}_{C_2H_2^+} \times \vec{p}_{CH_2^+}$) in the direction of the Z-axis, and $\vec{X} = \vec{Y} \times \vec{Z}$. The actual spatial orientation of the molecule within this fragment coordinate frame at the instant of photoionization remains, however, unknown. In larger systems, the fragments' emission directions are typically only loosely connected to the direction of the molecular bonds prior to the fragmentation, in particular if only few fragments are generated. Thus, the fragment frame (X,Y,Z) deduced from the ion direction measurement and a desired molecular coordinate frame (x',y',z') linked to its structure are typically skewed by some unknown angles ($R_X, R_Y$ and $R_Z$, i. e., the rotation angles with respect to the X, Y, and Z axis). In addition, the measured ion momentum vectors alone do not provide any information on the handedness of the ionized molecule, as they define a plane and thus leave the sign of the Z-axis open, as depicted in Figs. 1B and C. Both, this information and the information on the skew-angles are, however, encoded in the electron diffraction pattern.

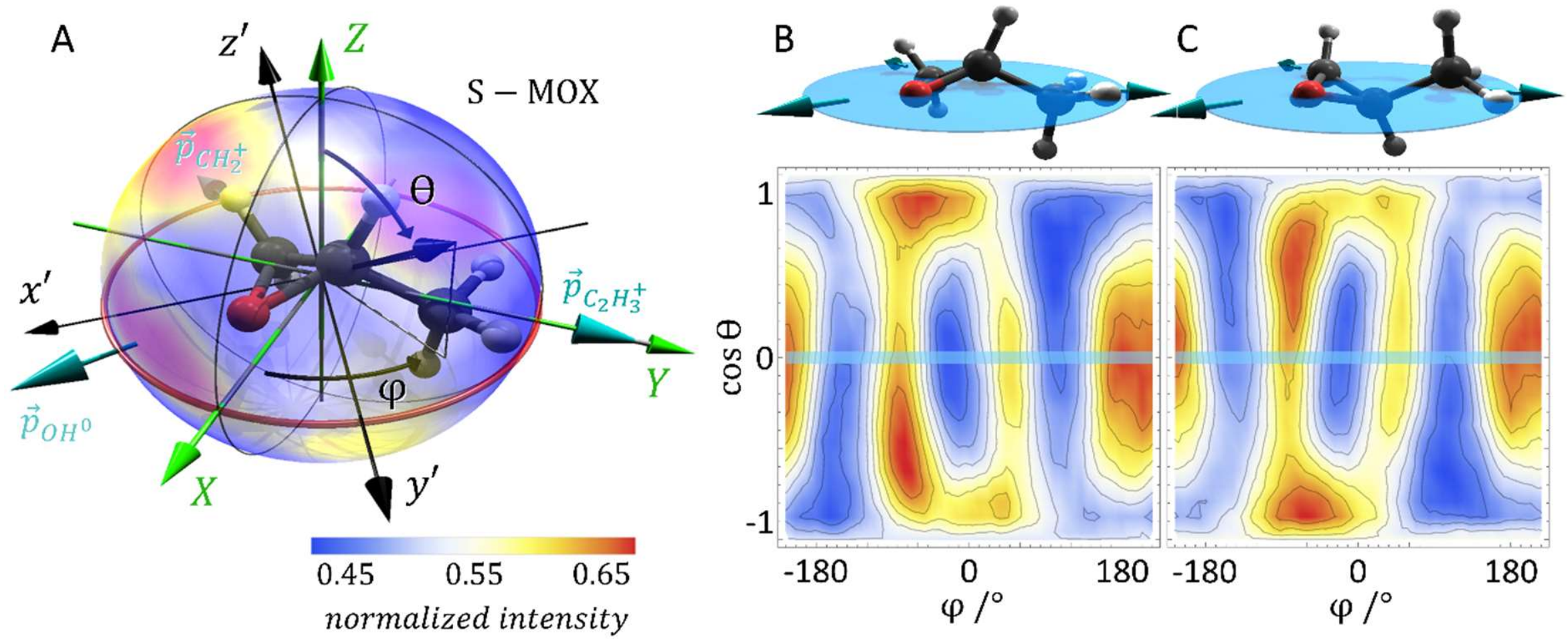

**Figure 1: Three-dimensional interference pattern of the scattered electron wave in the molecular frame of reference. A** Spherical representation and definition of the fragment (X,Y,Z) and molecular (x',y',z') coordinate frames. For larger molecules, the fragment emission directions (i.e., their momenta after Coulomb explosion) do typically not coincide with molecular features such as bonds. As a result, the molecular coordinate frame (x',y',z') at the instant of photoionization is rotated against the fragment frame (X,Y,Z). The panel depicts the methyloxirane molecule employed in our studies and the turquoise arrows show the directions of the measured momentum vectors of the fragments ($CH_2^+$, $C_2H_3^+$, and $OH^0$), which were used to generate the fragment (X,Y,Z) coordinate system, as discussed in the text. The electron wave employed for probing the molecular structure has been emitted from the oxygen 1s orbital. The surrounding-colored sphere shows the resulting three-dimensional probability distribution of the emission direction of the 11.5 eV photoelectron. The emission distributions are averaged over all incident directions of the ionizing light. **B** Same data as in **A** in a color-map representation. **C** is as **B** for the R-enantiomer. The mirror symmetry regarding the enantiomers is highlighted by the horizontal line to guide the eye at $cos\,\theta = 0$ in **B** and **C**. A visualization of the two enantiomers and their orientation in the molecular frame are given above B and C.

To extract the structural information from the experimental data we use the procedure that is outlined in Fig. 2. We start with an initial guess for the molecular structure and compute the photoelectron interference pattern in a guessed molecular frame (x',y',z'), which is assumed to coincide with the fragment frame. Then, we compare this pattern to the pattern obtained in our experiment, which is provided in the fragment frame (X,Y,Z). To quantify the agreement, we introduce the distance parameter $d_2$ between the renormalized experimental and computed interference pattern. This parameter depends, as well, on the relative rotation between (x',y',z') and (X,Y,Z) quantified by the rotation angles $R_X$, $R_Y$ and $R_Z$ and the guess of the handedness.

$$d_2 = \left(\iint \left(I_{Norm}^{Exp}(\varphi, \cos(\Theta)) - I_{Norm}^{Comp}(\varphi, \cos(\Theta), R_X, R_Y, R_Z)\right)^2 d\varphi d\cos\Theta\right)^{\frac{1}{2}}$$

We now determine the skew between the coordinate frames (x',y',z') and (X,Y,Z) varying the three rotation angles in order to obtain the smallest distance parameter $d_2$ (Fig. 2C).

The minimized value of $d_2$ (i.e., after applying the rotation) is then used to quantify the overall agreement between the measured and the computed interference pattern for the initially hypothesized molecular structure and handedness. This procedure is then repeated with a slightly adjusted molecular structure as an input in order to further minimize $d_2$. The model structure, which provides the smallest distance parameter $d_2$ is assumed to be responsible for the measured interference pattern, thus providing the molecular structure and coordinate system at the instant of ionization.

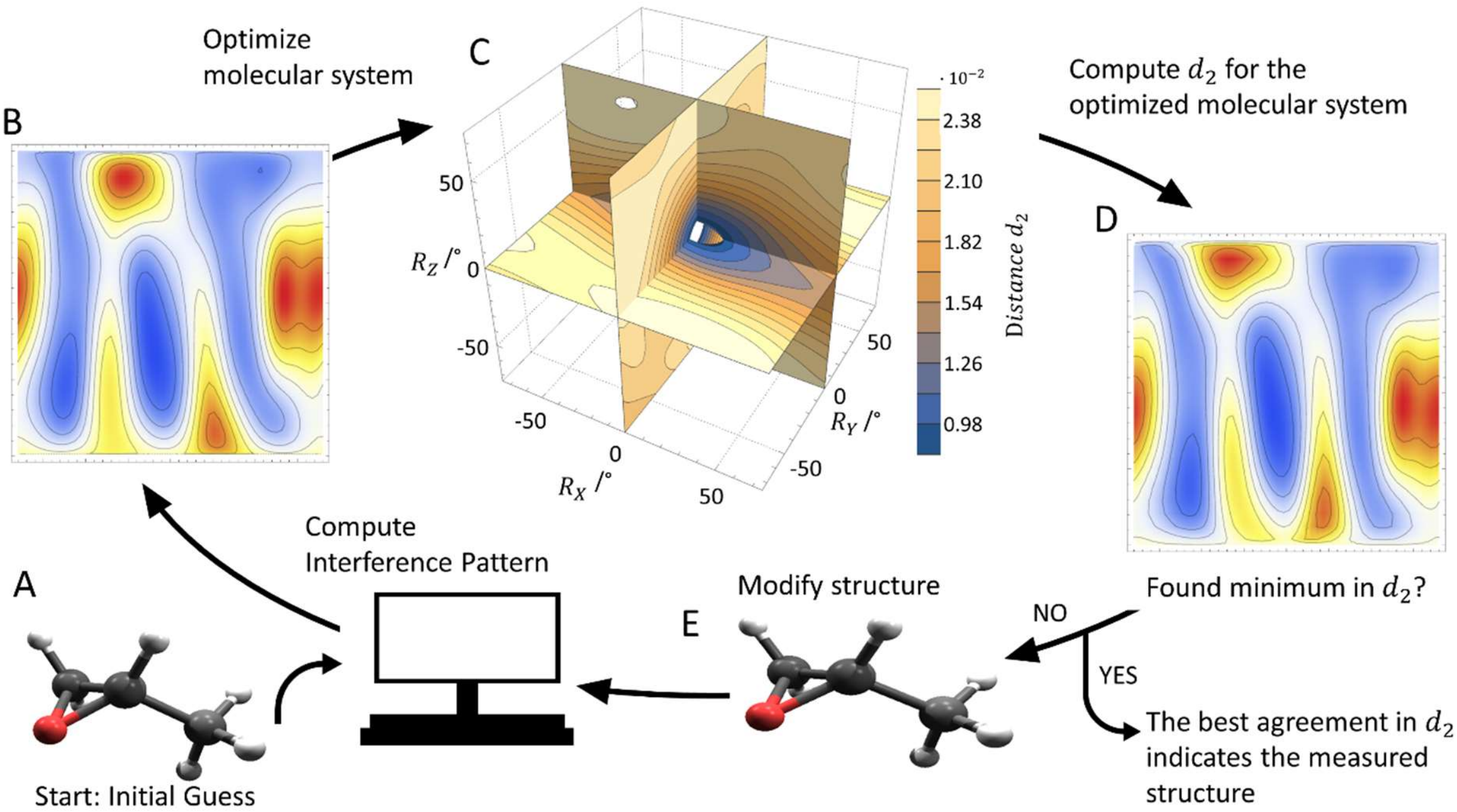

**Figure 2: Sketch of the optimization procedure for obtaining the molecular structure from the measured electron interference patterns.** Starting with an initial guess (A) the interference pattern of the photoelectron in a chosen molecular frame (x',y',z') is computed (B). This computed pattern is then compared to the measured interference pattern in the fragment frame (X,Y,Z). Typically, the fragment frame and molecular frame do not coincide. The skew between the two systems (given by the three rotation angles $R_X$, $R_Y$ and $R_Z$) is determined by finding the minimum value of distance parameter $d_2$ for the computed molecular structure (C). D shows the computed interference pattern from B in the rotated molecular frame (x',y',z'). After the rotation, the minimized value of $d_2$ is used to quantify the agreement between measured and computed interference pattern for a specific hypothesized molecular structure. The molecular structure is slightly modified, and the interference pattern is recomputed (E). The molecular structure and coordinate system at the instant of ionization are obtained for the best agreement between measured and computed interference patterns, i.e., for lowest $d_2$.

In more detail, in order to actually calculate the distance parameter $d_2$, we apply the following procedure: The minimum value occurring in the interference pattern is first subtracted from the pattern and then the pattern's integral is normalized to one. For each calculated molecular structure, the molecular system in coordinate space used in the calculation must be connected to the measured fragment momentum vectors. To do this, we rotate the measured and calculated interference pattern with respect to each other, applying the X-Y-Z convention (roll, pitch and yaw angle: $R_X$, $R_Y$ and $R_Z$). We determine the transformation that connects the fragment system defined by the measured ionic momenta and a molecular system used in the computation by searching for the minimum in $d_2$ in a scan over the yaw, pitch and roll angles (see Fig. 2). We scan all angles in steps of one degree. This step size is small enough to ensure that the residual error in the molecular frame does not influence the result presented in the following.

In order to test our approach, we employ the two enantiomers of methyloxirane as benchmark systems. As outlined in the experimental methods section, we are using a photoelectron of 11.5 eV kinetic energy emitted from the oxygen *K*-shell for the diffraction imaging and a naturally occurring subsequent Auger decay for the generation of two ionic and one neutral fragments. As shown in Figs. 1B and 1C, the observed interference pattern is vastly different for the two enantiomers making chiral discrimination straight forward. A comparison to the modelled pattern shown in Fig. 2D provides the information on the absolute configuration. As the geometry of methyloxirane is well known in the literature *(47)*, we employ the algorithm described above (and illustrate the high sensitivity of our approach) in order to extract the exact location of distinct atoms inside the molecule. A corresponding table of the atomic coordinates at its equilibrium can be found in the supplemental material of *(44)*. As a first example, we show in Figs. 3A-3F the effect of a modification of the $CC^* - O$ distance in the oxirane ring around the equilibrium structure. The resulting variation of $d_2$ is shown in Fig. 2F and implies, that we are sensitive to a change of 5 % of the geometry-optimized $CC^* - O$ distance. For comparison, a similar relative accuracy of a few percent for a bond length measurement has been demonstrated recently employing LIED examining OCS molecules *(48)*. Particularly challenging for other methods of structure determination is the assignment of the location of hydrogen atoms. Electron scattering is known to be sensitive also to such weak scatterers. Accordingly, in a second demonstration, we investigate the sensitivity of our approach to a change of the $C^* - H$ bond length. The corresponding results are depicted in Fig. 3G which confirms, that within 5 % discrepancy, the correct bond length between the chiral carbon atom and the adjacent proton attached to it, has been found via the smallest value of $d_2$.

The experimental statistical error is smaller than the plotted dot size; however, different sources of systematic errors might alter the exact value of $d_2$. The quality of the reconstruction as well as the achievable resolution depends on multiple factors, which cannot be easily quantified. For example, the recorded interference pattern has a finite experimental resolution. It is however not easy to estimate how resolving fine details of the interference pattern effects the reconstruction of geometrical features in the end. First estimates on the presented data suggest, that the experimental resolution is not limiting the accuracy of the geometrical reconstruction. Furthermore, background from other fragmentation channels or a complex interplay between areas of reduced detection efficiency on the electron and ion detectors might affect the geometry-reconstruction, as well. Yet, we expect that the statistic of the recorded datasets has a stronger impact on the reconstruction quality in our present study. An additional source of errors is connected to the approximation that is made in the ab initio computations. Each molecular configuration was calculated only for fixed internuclear distances (i.e., a single fixed molecular geometry) and only for a single photoelectron energy. In the experiment, however, the signal is averaged over a certain distribution of photoelectron energies and real-life molecules exhibit vibrational motion. In the case of a non-linear relationship between the influencing parameters and the observed variable, the mean value of the observed variable generally does not correspond to the mean value of the influencing factors. Therefore, an estimate of our resolution when determining the three-dimensional position of the atoms is provided by considering how close the minimum in $d_2$ comes to the result of the geometry-optimized structure when scanning across different molecular structures. Thus, the resolution is estimated to be of similar magnitude as our chosen step size of $\sim 6 \cdot 10^{-12}$ m. Please note, that with our technique the spatial resolution is not limited by the photoelectron's wavelength of 3.6 Å.

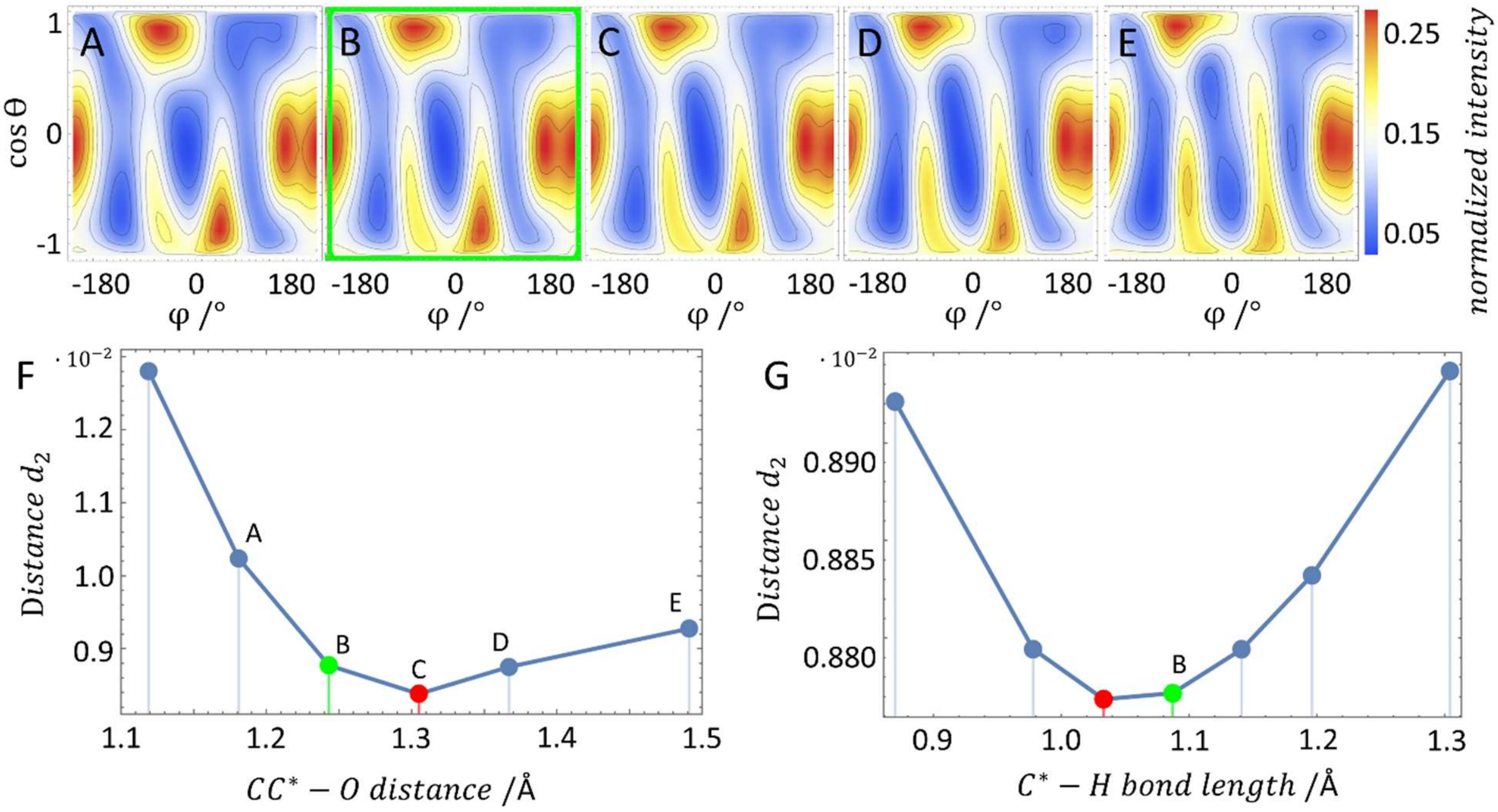

**Figure 3: Determination of the molecular structure via the best agreement between measured and computed interference pattern of the photoelectron. A–E** Interference patterns from a scan in which the $CC^* - O$ distance in the oxirane ring is set to 95, 100, 105, 110 and 120 % of the optimized structure. Our structure retrieval algorithm leads within 5 % accuracy to the energy-optimized structure highlighted in green (**F**). **G** A corresponding scan of the $C^* - H$ bond length demonstrates the sensitivity of the interference pattern to weak scatterers such as hydrogen. The smallest distance in $d_2$ leads, again, within 5 % accuracy to the $C^* - H$ bond length of the energy-optimized structure.

## Conclusion

Combining partial Coulomb explosion imaging of a large molecule with the measurement of the photoelectron diffraction pattern in the molecular frame and quantum chemical computation allows for precise structural analysis and chiral discrimination of molecules in the gas phase. Unlike established X-ray or several of the electron diffraction techniques, our approach does not require a molecular crystal. Contrary to traditional Coulomb Explosion Imaging, it is scalable, so that larger molecules can be examined, as well. The only requirement is that the molecule breaks sufficiently fast into at least three fragments (of which at least two are charged), so that the measured ion momentum vectors are linked to the molecular orientation at the time of ionization. In addition, by adjusting the photon energy, distinct atoms of the molecule can be addressed and the emission source of the probing electron wave inside the molecule can be selected. By applying pump-probe schemes, the method will allow for tracking changes in the molecular structure on a femtosecond time scale *(49,50)* in the future. The described approach is in principle general and can be extended to larger molecules. If necessary, required calculations can rely on density function theory (DFT) based methods which are usually not limited by molecular size, like e. g. the TDDFT B-spline LCAO formalism *(51).*

## Data availability

All data needed to evaluate the conclusions in the paper are present in the paper. Additional data related to this paper may be requested from the authors. Correspondence and requests for materials should be addressed to K.F. (fehre@atom.uni-frankfurt.de), P.V.D. (demekhin@physik.uni-kassel.de), or M.S.S. (schoeffler@atom.uni-frankfurt.de).

## Author Contributions

The experiment was conceived by M.S.S. and R.D. The experiment was prepared and carried out by S.G., G.K., S.E., F.T., J.R., A.H., D.T., C.J., M.P., S.Z., F.W., M.W., M.K., M.H, L.Ph.H.S., A.K., A.H., L.B.L., A.E., R.B., H.F., K.U., T.J., J.B.W., and M.S.S. Data analysis was performed by K.F. and M.S.S. Theoretical calculations were performed by P.V.D and N.M.N. All authors discussed the results and commented on the manuscript. K.F., P.V.D., R.D., T.J, and M.S.S. wrote the paper.

## Conflicts of interest

The authors declare that they have no competing interests.

## Acknowledgements

We acknowledge synchrotron SOLEIL (Saint-Aubin, France) for the provision of experimental facilities. The experiment was carried out at the beamline Sextants. We thank D. Reich for discussion. We thank A. Czasch und O. Jagutzki from Roentdek GmbH for support with the detectors. We thank the staff of SOLEIL for running the facility and providing beamtimes under project 20140056 and 20141178 and especially SEXTANTS beamline for their excellent support.
This work was funded by the Deutsche Forschungsgemeinschaft (DFG) – Project No. 328961117 – SFB 1319 ELCH (Extreme light for sensing and driving molecular chirality).

K.F. and A.H. acknowledge support by the German National Merit Foundation. M.S.S. thanks the Adolf-Messer Foundation for financial support. H.F. and K.U. acknowledge the XFEL Priority Strategy Program of MEXT, the Research Program of “Dynamic Alliance for Open Innovations Bridging Human, Environment and Materials”, and the IMRAM project for support.